# Glyph: Visualization Tool for Understanding Problem Solving Strategies in Puzzle Games

Truong-Huy Dinh Nguyen, Magy Seif El-Nasr, Alessandro Canossa
PLAIT Lab, College of Arts, Media, and Design, Northeastern University, Boston, MA 02115
{tru.nguyen, m.seifel-nasr, a.canossa}@neu.edu

## ABSTRACT
Understanding player strategies is a key question when analyzing player behavior both for academic researchers and industry practitioners. For game designers and game user researchers, it is important to gauge the distance between intended strategies and emergent strategies; this comparison allows identification of glitches or undesirable behaviors. For academic researchers using games for serious purposes such as education, the strategies adopted by players are indicative of their cognitive progress in relation to serious goals, such as learning process. Current techniques and systems created to address these needs present a few drawbacks. Qualitative methods are difficult to scale upwards to include large number of players and are prone to subjective biases. Other approaches such as visualization and analytical tools are either designed to provide an aggregated overview of the data, losing the nuances of individual player behaviors, or, in the attempt of accounting for individual behavior, are not specifically designed to reduce the visual cognitive load. In this work, we propose a novel visualization technique that specifically addresses the tasks of comparing behavior sequences in order to capture an overview of the strategies enacted by players and at the same time examine individual player behaviors to identify differences and outliers. This approach allows users to form hypotheses about player strategies and verify them. We demonstrate the effectiveness of the technique through a case study: utilizing a prototype system to investigate data collected from a commercial educational puzzle game. While the prototype's usability can be improved, initial testing results show that core features of the system proved useful to our potential users for understanding player strategies.

## Categories and Subject Descriptors
H.5.m [Information Interfaces and Presentation (e.g., HCI)]: Miscellaneous

## 1. INTRODUCTION
Understanding player strategies is an important question when analyzing player behavior, both for academic researchers and industry practitioners. To game designers, investigating players' strategies and behaviors over time informs their understanding of how well the game mechanics and pacing works and allows them to uncover design issues. To academic researchers using games for serious purposes such as education, strategies adopted by players are indicative of their cognitive progress in relation to the goals, such as learning. This paper describes a novel system that allows users to interact with visualized gameplay behavioral data to gain better understanding of players' problem solving strategies.

Traditionally, there are two main approaches to analyzing player behavior in games. The first includes the use of one or more of the following techniques: focus groups, think-aloud protocols, in-lab play-testing, physiological recordings and retrospective interviews [4, 12, 15, 27]. All of these techniques require direct interaction with participants. For example, physiological assessments require users to be in the lab with sensors, play-testing requires users to be in a lab playing a game while researchers observe or record their session. Additionally, all these techniques do not scale with the number of subjects, since they require the subjects to be physically present during the course of the study. Most of these techniques are also qualitative and thus are prone to subjective opinions, which can be biased.

Another approach to investigating players' problem solving strategies is through the use of telemetry, i.e. logs of players' moment-to-moment in-game actions [27]. Although telemetry does not collect affective states, its process is automated, allowing data to be reliably collected in an easily scalable manner. Using this data, analysts can then use quantitative techniques leveraging machine learning and data mining to process or analyze the data. Analysts have also been using visualization techniques to gain better understanding of the data and also help designers and other stakeholders gain insights from the data analysis results [1, 2, 5, 7, 8, 10, 14, 17, 19–22, 26, 28, 30].

In order to understand strategies enacted by players, it is important for stakeholders to interact directly with behavior data as immediate feedback is crucial for formulating and evaluating hypotheses. The matter is complicated because behavior data can be aggregated to provide an overview of possible strategies but can also be treated as individual play traces to identify differences and similarities between individual players. Therefore, stakeholders need to be able to directly manipulate both aggregated behaviors and individual play traces, receiving immediate feedback to expedite the process of formulating and evaluating hypotheses.

Visual analytics has been in the forefront of promising techniques to unpack temporal data, such as gameplay behaviors, to answer questions pertaining to problem solving strategies and behavioral patterns. Current visual analytics and visualization systems allow for either analysis of individual behaviors or examination of aggregated behaviors [1, 2, 5, 7, 8, 10, 14, 17, 19–22, 26, 28, 30]. To the authors' knowledge, there are no visual analytics systems capable of handling large numbers of players' temporal behavioral patterns, providing an overview of the whole community and at the same time allowing investigation of individual players' patterns. Furthermore, current systems do not abstract behavioral patterns, while at the same time allowing quick views of unique and popular patterns. For example, most of the current visualization methods provide a single main view of the data, with information delivered to users either as overlays or as sidebars. The task of behavior comparison is often delegated to users. This process can be taxing, overloading their visual cognitive process over time, which can reportedly reduce the quality and productivity of the investigation process [16, 25].

To tackle these limitations, we propose a novel interactive visualization system. In particular, we propose a multi-view interactive visualization technique that makes it easy for users to both gain an overview of possible behaviors and also allows close examination of individual play traces, comparing behavior traces and identifying relevant similarities and differences, without any additional mental processing. In one view, the whole population of traces is displayed using node-link representation with nodes corresponding to states and links to actions; this will be referred to as 'state view'. Simultaneously in the other view, each behavior sequence present in the population is encoded as a single node; distances between them represent their similarity while size represents their popularity; this will be referred to as 'sequence view'. With synchronized information highlighting, users can quickly identify differences between action sequences in terms of both moment-to-moment details (in state view) and as a whole (in sequence view). By sharing the amount of visual information displayed to users among the views, each view provides concrete information pertaining to specific questions, thus not overloading users' cognitive process when interacting with the system. A notable advantage of our approach that makes it generalizable to many game genres is that it adopts an action-based state comparison scheme, a domain-independent metric to compare behavior sequences, instead of requiring domain experts to furnish a dissimilarity function.

In this paper, we first present related works on other approaches in aiding users to understand player strategies. Next, we describe a phase of requirement analysis that we undertook to understand the limitations of current solutions. Subsequently, we will discuss a case study used to illustrate the use of the visualization system. The case study is a game called Wuzzit Trouble (BrainQuake, 2013), a commercial educational puzzle game that was developed to teach students about algebraic mathematics. We will then explain in detail different components and design decisions of the prototype visual analytics system. Finally, we will discuss our initial testing results of the prototype illustrating both the features and limitations of the current system, leveraging feedback from potential stakeholders.

## 2. RELATED WORKS

In order to understand play strategies, stakeholders often adopt one of two main approaches: lab studies [4, 11, 12, 15], such as playtesting, or remotely tracking gameplay logs while users interact with the game in their naturalistic setting [27].

The first includes the use of one or more of the following techniques: focus groups, think-aloud protocols, in-lab play-testing, physiological recordings, and retrospective interviews [4, 11, 12, 15]. All of these techniques require direct interaction with participants. They therefore do not scale well with the number of subjects, since they require the subjects to be physically present during the course of the study. Besides, they are prone to subjective opinions, which can be biased, and do not provide moment-to-moment action records.

The alternative option is the use of game telemetry [27] and game analytics – a set of techniques designed for collecting and analyzing play traces, i.e. records of players' in-game actions and states when they engage with the game. Thanks to game analytics methods the process of collecting player behavior data is automated and easily scalable [27]. Quantitative techniques leveraging machine learning and data mining can then be adopted for processing the data.

Since such analyses are textual or numerical in nature, there has been an emerging trend of adopting visualization tools to present the analysis results in an easily interpretable form. Abstractly, there are four main types of visualization techniques: simple bar charts [14, 19, 20, 22, 26], heatmaps [2, 7, 8], movement visualizations [5, 10, 21], and node-link graphs [1, 17, 28, 30].

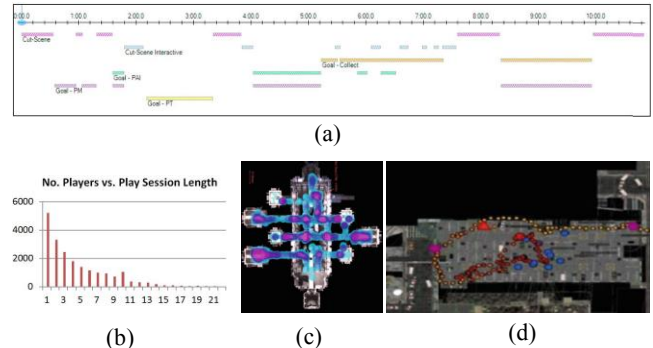

**Figure 1. Visualization techniques; (a) chart showing a player's behavior pattern, (b) chart showing aggregated statistics, (c) heatmap, (d) movement visualization. Images (a, c, d) are reproduced with permission from respective authors [20, 27]**

Simple bar charts present users with visualizations of aggregated statistics on the whole set or some user-defined subsets of players [14, 19], or behavior patterns of specific players in some temporal order [20, 22, 26]. Popular aggregated statistics include kill/death ratios, experience points gained, daily numbers of play rounds, etc. [19]. For instance, Figure 1b shows the bar graph of the number of players charted against session length in a puzzle game; most players only play at most 10 levels within one session. Behavior patterns, on the other hand, are projections of player in-game behaviors onto some high-level semantic space, e.g., movement actions mapped as path target pattern, or item pickup as collection pattern [20]. By encoding different patterns with different colors and placing them at different height, a player's behavior trace can be visualized as shown in Figure 1a.

Heatmaps [2, 7, 8] and movement visualizations [5, 10, 21] are techniques used to analyze spatial behaviors. They tie data points to their respective geographical locations on the game map and display pertinent information using visual cues such as color clouds, color-coded icons, or simple bar charts. For example, Figure 1c shows a heatmap of player death events in a role playing game, while Figure 1d depicts the movement visualization of a player, with circles representing location snapshots and temporally color-coded (yellow earlier, red later). Heatmaps excel in showing population behaviors, but make individual play trace comparison an involved process. At the same time it is impossible to evaluate behaviors unfolding over time. Movement visualizations, on the other hand, focus on showing individual traces, leaving questions on aggregated behaviors open to users.

Node-link representations visualize high-dimensional gameplay data that do not naturally map to a 2D map, such as those collected from puzzle games [1, 17, 28, 30].

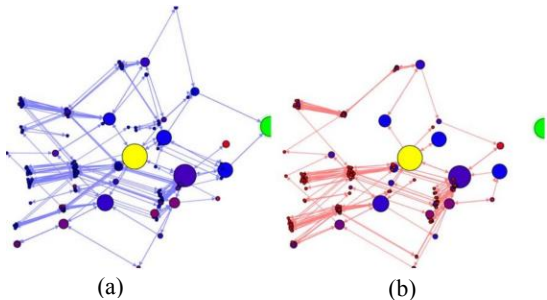

**Figure 2. Playtracer visualizations of (a) winners and (b) losers in the same game level**

Playtracer [1] is a visualization system that adopts node-link representation approach. Gameplay data is abstracted in to a directed network, with nodes encoding game states and directed links actions that transform from one state to another (Figure 2). The placement of nodes is determined using multidimensional scaling, while node size indicates popularity and color terminal status (starting node is yellow, ending node green). Multidimensional scaling is a technique that takes as input a dissimilarity function (or matrix) for each pair of data points and returns a projection of the high-dimensional data set to a 2D coordinate system [23]. Selecting an appropriate dissimilarity function for a specific game domain is a challenge, so as not to bias analysis. Playtracer captures aggregated population behaviors visually, i.e. common patterns can be detected through clusters of links and nodes. However, it is not trivial to answer questions related to uniqueness and commonality of patterns, such as "What are the top 10 most popular sequences?", or "How is the most frequent losing pattern compared to winning patterns?" Besides, comparing individual patterns requires users to simultaneously examine multiple windows, each displaying a single trace. It could be manageable to compare two traces, but comparing ten traces is not a trivial task. Other similar node-link visualization techniques are proposed [17, 28, 30] but none addresses these limitations.

As mentioned in Section 1, existing approaches are limited either because of subjective biases and poor scalability or because they focus either on individual traces or overall aggregation. The system we are proposing here is based on objective data, scalable, and provides a synchronized interactive visualization of both aggregated overview and individual traces.

## 3. CASE STUDY – WUZZIT TROUBLE

*Wuzzit Trouble* is a commercial game, released as a free download by the startup company BrainQuake in Fall 2013 [6]. The game is designed to provide arithmetic-based puzzles with increasing difficulty in a fashion that circumvents the usual symbolic notation of arithmetic in an effort to break the symbol barrier, a widely known obstacle in arithmetic problem solving [6]. The goal of the game is to free creatures called Wuzzits from their traps by collecting all the keys in a level. The keys hang on to a large wheel that can be rotated; players collect a key when the marker at the top aligns with the position of that key.

For instance, in Figure 3, the marker is at number '0' and needs to be moved to number '19' and also to number '51' to obtain both the keys needed to free the trapped Wuzzit. To align the marker with the said number, players rotate the large wheel clock-wise or anti-clockwise, by turning the cogs provided. The distance or number of units moved by the large wheel depends on the cog that is used (Figure 3). For example, if a player turns cog 3 clockwise twice, the marker moves 6 pegs to the right; or if the player turns cog 13 anti-clockwise once, the marker moves 13 pegs to the left.

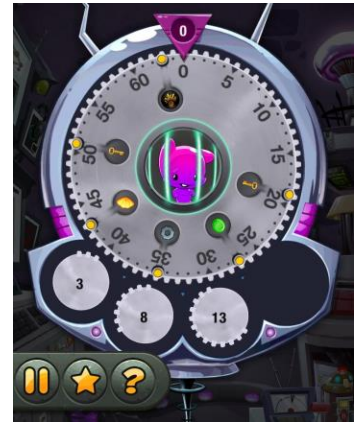

**Figure 3. Wuzzit Trouble, Stage 2, Level 3, three cogs provided with 3, 8, and 13 teeth**

Each cog can be turned up to five times in one move to generate a five-step turns of the wheel, offering up to five opportunities to collect a key (or other item) with a single move. This is a critical gameplay mechanic to learn in order to free the Wuzzit with the smallest number of cog rotations, and in doing that the player can beat the level gaining the most number of stars. In addition, the pegs with unique gems attached to them, like the green gem between 25 and 30, rewards the player with bonus points. At higher levels, there are items that deduct points from the players, so for players who value high points, they should avoid these items as much as possible. The overall score at each level is determined by the number of keys and bonus items collected.

## 4. REQUIREMENT ANALYSIS

So what do researchers want to know about players' strategies? A requirement analysis study was conducted with our targeted users, including two learning scientists, via the means of informal interviews; each session lasted 30-60 minutes. We asked several questions pertaining to their goals and use of the visualization system to unpack learning objectives and behaviors. Two example questions are as follows:

1. What are the goals of your analysis in Wuzzit Trouble?
2. What features would you like to have with this game?

The results of this study allowed us to understand what the learning scientists wanted from a data analysis system. We found that they are interested in investigating strategies and patterns of problem solving behaviors. They expressed a demand to have a tool that will allow them to examine and compare action paths, which helps (1) recognize common behaviors in groups of different backgrounds, as well as (2) identify outlier performances.

The first capability allows them to validate design decisions, which are driven by education goals, intended progress, and milestones, from actual player behaviors. For instance, if a designer anticipates different solutions for a level, but players' solutions do not show enough diversity, the design may need some tweaking to make the different pathways through the problem solving exercise clearer. The second capability helps educators identify kids who might need additional attention, due to their extraordinary ability to reach education goals in abnormal, either negative or positive, ways. This is extremely useful for class instructors.

Using these results as requirements for our system, we then developed a visual analytics system called Glyph. We describe Glyph in the next sections.

## 5. VISUAL REPRESENTATIONS IN GLYPH

With the focus of visualizing action paths and providing means to compare them, we adopt two visual representations: a state graph and a sequence graph. The state graph shows all action paths in a single view as a node-link diagram, while the sequence graph encodes action sequences as nodes, the distance of which provides a visual representation of how dissimilar corresponding action sequences are.

### 5.1 State Graph

To understand strategies, we abstract the current collected game behavioral data into a state space graph diagram, in the form of a node-link directed network.

In essence, following Juul's convention [13], we represent **nodes** in the graph as different game states and **directed links** as actions that users took to get from one state to another. Thus, as users interact within the game world and solve problems, they will be navigating within the state graph represented in this visualization. Specifically, a game state captures all current information associated with in-game entities; for instance in Wuzzit Trouble, current marker position, and the pickup statuses of keys and items make up a state's information. A player's action, such as turning a specific cog a number of times, leads to an update to the game state, in which player's current marker position is changed, as well as the item statuses should one or more are picked up.

Figure 4a depicts an example state diagram. The size of the state nodes and thickness of links indicate the popularity of corresponding states and actions in the data set, i.e. the larger the nodes and thicker the links means more players traversed them. One blue node is the initial starting state, multiple red nodes possible end states (in Wuzzit there could be multiple end states), and yellow nodes mid-states.

The positions and distances between nodes are determined using force-directed placement [5]. The layout algorithm simulates the physical process of attracting particles on a force field, placing nodes that are highly interconnected close by and nodes less connected apart. In our visualization, groups of highly connected nodes indicate similar behavior patterns, comprising of actions and states often co-existing in the same patterns. For instance, in Figure 4a, the cluster of large nodes and thick links in the top right corner represents a group of highly popular behavior patterns, whereas different clusters in Figure 4b indicate groups of similar patterns that ending in the same red end states.

This type of diagrams, while informative, can get complicated and cluttered quickly especially with complex levels and many users, see Figure 4b. The figure depicts data from hundreds of users. Although the state graph exhibits some clustering of state nodes, tracking and comparing multiple play traces can be hard. Edge bundling [18] would simplify the graph but would also render impossible any form of analysis on individual strategies.

### 5.2 Sequence Graph

Sequence graph view visually shows the popularity and similarity of sequence patterns exhibited by users (Figure 5). Each node in the graph represents a full sequence, labeled by their popularity rank, i.e., the smaller the number, the more popular the respective sequence is, with 0 being the most popular sequence. Distance between nodes is a measure of how different they are, computed using Dynamic Time Warping (DTW) [3]. DTW is a technique often used in speech recognition, computing the numeric difference of sequences of unequal lengths. In order to do so, DTW non-linearly warps the involved sequences to find the optimal way to match between-sequence items, which minimizes the total difference. As such, to use DTW for computing sequence dissimilarity, we need furnish as input a metric function to compare game states.

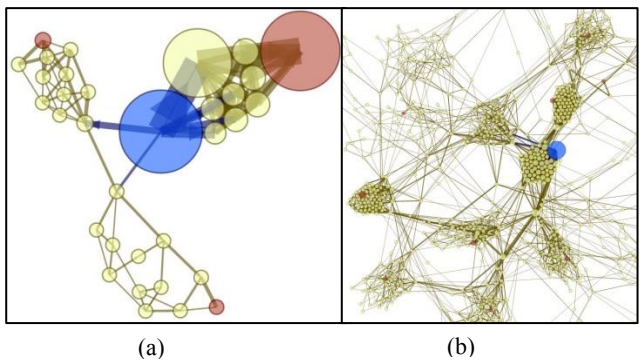

(a)   (b)

**Figure 4. State graph view of Wuzzit behavior data in two sample levels; (a) is simpler than (b); starting node is blue, end node red, and transition nodes yellow.**

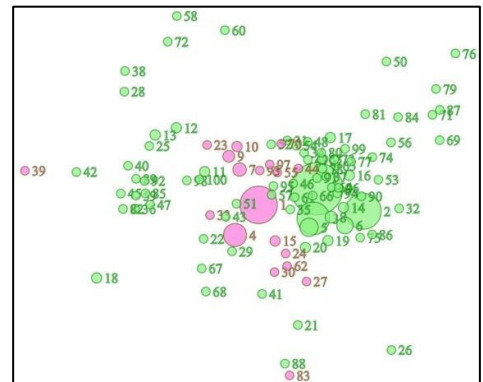

**Figure 5. Sequence graph; green nodes are completed sequence, pink nodes uncompleted. Node labels indicate sequence popularity; the smaller the more popular.**

**State difference:** We define the difference $d(s_1, s_2)$ of state $s_1$ and $s_2$ as the smallest number of actions needed to transform from $s_1$ to $s_2$ or vice versa. If there is no way to transform from one state to the other, $d(s_1, s_2)$ is set to infinity.

The more actions required to transform the states, the more different they are. Note that this definition accounts for the case when one state can be transformed to the other but not vice versa, such as when $s_1$ is one with all keys collected and $s_2$ not in Wuzzit Trouble. In this case, $s_1$ cannot be transformed to $s_2$, but $s_2$ can; thus the difference of $s_1$ and $s_2$ is the smallest number of actions to transform $s_2$ to $s_1$. This domain-independent metric is applicable to in all state-oriented situations, i.e. players engage with the game to achieve some desirable game state, regardless of the action sequence they selected to reach there. Given this general notion of state difference, DTW algorithm computes sequence difference using dynamic programming.

**Dynamic Time Warping:** Given $d(s_1, s_2)$ as the difference of any state pair $s_1$ and $s_2$, the difference $D(a, b)$ of two sequences $a = \{s_1, s_2, \ldots, s_n\}$ and $b = \{q_1, q_2, \ldots, q_m\}$ is computed as $D(n, m)$ as follows

1. Initialization:
   a. $D(0,0) = 0$
   b. For $i$ in $[1, n]$ and $j$ in $[1, m]$:
   $$D(i, 0) = D(0, j) = inf$$
2. Recursion: For $i = 1$ to $n$ and $j = 1$ to $m$
$$D(i,j) = d(s_i, q_j) + min \begin{bmatrix} D(i-1, j), \\ D(i, j-1), \\ D(i-1, j-1) \end{bmatrix}$$
3. Return $D(n, m)$

In the initialization step, $D(0,0)$ represents the distance of empty sequences, thus valued as 0. $D(i, 0)$ and $D(0, j)$, i.e. the distance between an empty sequence and a non-empty one, are set to infinite value, indicating that non-empty sequences are inherently different from empty ones. In practice, we replace the infinite distance value by a large number that exceeds the length of all possible sequences. In the recursion step, the distance value is the smallest among three alternatives, insertion, i.e., $D(i-1, j)$, deletion ($D(i, j-1)$), and match ($D(i-1, j-1)$). As such, DTW can be treated as a generalization of minimum edit distance [29], in which state comparison metric replaces simple matching operations.

When the sequence distance values are encoded as link length in the force-directed graph, clustering appears naturally, as similar sequence nodes clump together. As such, sequence nodes farther away indicate that the behaviors exhibited are different. The size of sequence nodes indicates the popularity of that sequence. Note that this approach of encoding node dissimilarity as link lengths in a force-directed graph is equivalent to a metric multidimensional scaling layout [23]. As such, other multidimensional scaling techniques can be adopted as well should they be deemed domain-specifically more appropriate.

Figure 5 shows a sample sequence graph. Green nodes depict complete sequences, i.e. those that complete at some end state, and pink nodes incomplete sequences, i.e. those that players quit half way through the game. The graph shows many pink nodes mixed up with green nodes. This means while there are many complete sequences, there are a significant number of players that quit the level in similar manners, while it seems that they can still complete it. Such information helps notify researchers of subtle and delicate, albeit possibly detrimental, issues that can pass by without notice otherwise.

## 6. INTERFACE DESIGN AND VISUAL ENCODING IN GLYPH

The prototype system sports a dual-view interface that displays data in both state and sequence views (Figure 6a). Users interact with the system by (1) entering queries into input text boxes, and (2) mouse controls in the views. The system conveys query results to users as synchronous highlighted elements in the views, accompanied by text info displayed at the top of the view (State Node and Action Link info in Figure 6b and Sequence Node info in Figure 6c). Besides, details on the layout of the level are also provided as context information (Level Info in Figure 6b).

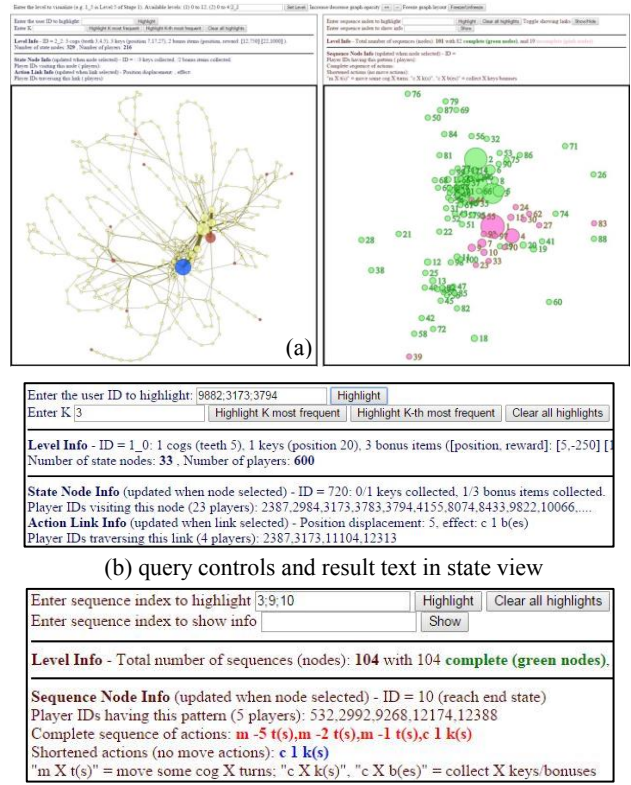

Figure 6. The main interface showing (a) state view (left) and sequence view (right). On the top of each view are query controls and result texts, (b) and (c).

### 6.1 Interaction Options

The system supports natural mouse controls, and provides two areas of text inputs as navigation and interaction options.

More specifically, within each view, users can drag and drop nodes to override the automatic layout mechanism and affix the nodes to desired locations. In densely populated graphs, this operation allows users to freely dissect node clusters for better investigation. Besides, users can zoom in and pan to regions of interest in the graphs for better view. Clicking on nodes and links updates node and link text info (Figure 6b and 6c), as well as getting them highlighted (see Section 6.2 for more details)

To query one or more individual sequences, users can key in respective user IDs (Figure 6b shows ID 9882, 3173, and 3794 keyed in). To query most frequent sequences, users can key in K, and get K most popular or just K-th most popular sequence (Figure 6b shows K=3). To compare different frequent patterns, users key in sequence IDs, for example, 3, 9, 10 (Figure 6c).

When a sequence is highlighted, its details are shown in both raw and condensed form (Sequence Node Info in Figure 6c). Condensed information only contains the part of the action sequences where meaningful actions are executed. For example, in Wuzzit Trouble, the meaningful actions are those that lead to the collection of items (keys or bonuses). As such, if a raw action sequence goes as "Move counter-clockwise 5 step, move counter-clockwise 2 step, move counter-clockwise 1 step, collect 1 key", its condensed form is "Collect 1 key" (Figure 6c). When these are displayed side by side, users are able to quickly switch between seeing the full form action sequences and just the important actions that might have been the motivations of the surrounding movement actions. In other games, we suspect that actions can be similarly classified depending on the specific domain. For example, in role playing games, there could be a lot of navigation and movement actions, but actions that really contribute to the play experiences are those involving engagement with in-game entities such as conversing with non-player characters.

## 6.2 Synchronized Sequence Highlighting

We adopt a synchronous information visualization approach to present query results to users. In particular, within the system, as users query about a sequence, both the sequence node and its respective behavior sequence in its full form are visually highlighted in the views.

Figures 7 shows three snapshots of the system, demonstrating different acts a specific researcher can execute using the system and the visual results. Figure 7a shows the graph with no interaction. Figure 7b shows the state and sequence graph resulting from the researcher selecting the most popular pattern, shown as big red circle in the sequence graph. In the state graph, the corresponding sequence of actions within the state space, thus the actual action pattern, is highlighted in the similarly red color. Figure 7c shows two views where the researcher selected an assortment of nodes in the sequence graph (right image); the state graph (left image) shows the corresponding paths through the state space, illustrating how players took different actions to go from start state (in light transparent blue) to end state (in light transparent red).

Figure 7 also demonstrates how different popular strategies can be observed quickly from the views. Recall that the sequence graph arranges the nodes based on similarity. Thus, for example in Figure 7, we can see that there are three clusters of different behavioral patterns, with the distance showing the level of similarity. By selecting one or many different sequences for highlighting, users can visually compare and contrast different player behavior. Note that corresponding to each sequence is a text label that shows the sequence of constituting actions. Specifically, the sequences highlighted in Figure 7c correspond to:

- SequenceID 0: Move counter-clockwise 5 steps, collect 1 key
- SequenceID 3: Move counter-clockwise 1 step, move counter-clockwise 1 step, move counter-clockwise 1 step, move counter-clockwise 1 step, move counter-clockwise 1 step, move counter-clockwise 1 step, collect 1 key
- SequenceID 15: Collect 1 bonus item, move counter-clockwise 5 steps, collect 1 key
- SequenceID 16: Collect 2 bonus items, move counter-clockwise 5 steps, move counter-clockwise 5 steps, collect 1 key

Upon comparing sequences, from this textual information, it shows clearly that the majority of players, captured in Cluster 1, go to the key, collect it, and move on to the next levels. In actualizing this strategy, some players demonstrate less optimal movement behavior (corresponding to SequenceId 3), in which they only move one step at a time to get close to the key before collecting it. A small number of players in Cluster 2 adopt a different strategy, moving to collect one bonus item before collecting the key for level completion. Lastly, a few players collect all bonus items in this level before completing it. This demonstrates different approaches that players can adopt when solving this particular level in Wuzzit Trouble, and various ways to manifest them. By examining the differences in behavior sequences, researchers/game designers can form hypotheses about how different player types approach this game level, and sometimes in broader sense, the intended arithmetic equations encoded in this level.

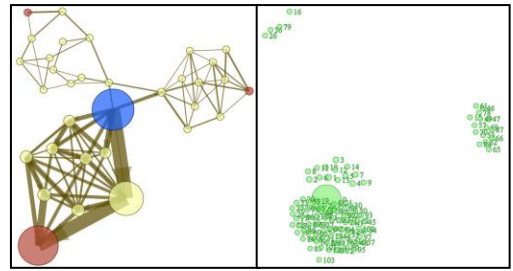

(a) No highlight

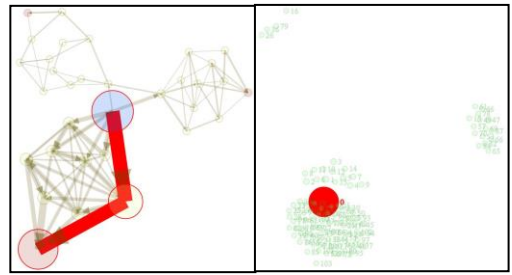

(b) Most popular play-trace

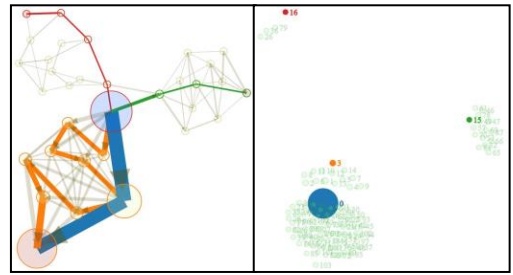

(c) Play-trace comparison

**Figure 7. Synchronized highlighting between two views allows quick inspection and comparison of play traces. The same sequences are highlighted with the same color in two views**

## 7. INITIAL TESTING OF GLYPH
## 7.1 Prototype Implementation

The system prototype uses D3.js, which leverages Canvas on modern web-browsers to display the views. As such, the system can be distributed on the web, allowing remote access and collaboration.

## 7.2 Setup
We invited several expert users to use the system for their own purposes in understanding player behaviors in Wuzzit Trouble. Two were learning scientists; they were interested in making sense out of player's in-gram behavior in relation to their arithmetic ability. As such, it is important to them to identify what the behavior norms are, as well as how a certain individual trace deviates from these norms, which helps them to form hypotheses on the "why".

The data used are collected from one and a half months long of play sessions, which consist of 100K level completion events. The data is segmented into subsets, each containing only play traces in a specific level. Within the tool, users can specify the level they would like to investigate, the corresponding data set of which is promptly loaded into the views.

We provided users a 2-page instruction sheet, which describes shortly the views and the kind of information they convey, as well as available interaction controls (mouse, keyboard). The tool is hosted online, so users can access and evaluate it from afar. After users have finished their interaction with the tool, we conducted a post-interview with them.

## 7.3 Results
The post-interviews demonstrate the advantages of our approach in the form of positive feedback, while informing us of room for improvement as critiques, as well as exposing some usability issues with the current prototype. While it is worth noting that this initial phase of testing was greatly affected by the usability of the interface, it is encouraging to us that despite the reported issues, many of our design decisions are highly appreciated by potential users.

### 7.3.1 Goals
Given no predefined task to carry out with the tool, users were achieving two different goals:

1. Understand the data and the traces
2. Explore different game level data

### 7.3.2 Positive feedback
The learning scientists rated the synchronous highlighting feature very highly. They also liked the incorporated analysis results on most frequent patterns. In particular, the ability to "select the highest x-th, top x highest, etc. and have those show up in both diagrams" helped them understand the common behaviors in each level, while one commented that the latter feature "gives users initial ways to begin their analysis." Besides, although primitive, textual information of the highlighted sequences and states was praised for letting users know the meaning of the actions.

As expressed by the scientist from BrainQuake, it is essential for them to be "able to follow the user's solution path to a problem and compare it with that followed by others", and that "Glyph seems perfectly suited to meet that need."

### 7.3.3 Critique
The interface had many limitations. First, users wanted to see a visual representation of the state. In the current prototype, we provided a text description of each level, informing users of the cogs available, the number of keys and their positions on the wheel, as well as the list of bonus items. An image showing this information instead would make it a lot easier to understand the puzzle at hand and thus unpack players' thinking when observing their moves. Additionally, users also asked for more details on the text information of the sequences.

In addition, there was some feedback concerning labeling and exporting data. Specifically, it was requested that we add the ability to label specific states or state sequences with keywords, and then export only the subset of the data that contain sequences with certain keywords for further investigation. For example, with players that make a lot of "move 1 step" in their sequences, users would like to flag those action sequences as "still have troubles with game rules". Later all, the players with this flag can be tracked for evaluation on their progress. Other keywords to tag other behavior patterns can be used to indicate different groups of players or behaviors that would be interesting to follow up.

While the system had some limitations, we believe the main feature and contribution of this system over previous work was clearly a win for the users who interacted with this system. We are currently planning for more testing sessions to gain additional feedback to revise the system and develop it to its full potential.

## 8. CONCLUSION
In this paper, we described our proposed visual analytics system Glyph, which contains features that allow easy understanding and investigation of player behavior and strategy. The system consists of two different views of the same data set: state view and sequence view. The two views provide different perspective about the same action sequences, thus allowing users to quickly find the answers to questions on both common patterns as well as individual moment-to-moment behaviors. A prototype of the technique was developed and evaluated. While the usability of the prototype is still considered problematic, the main features are highly rated by participants, with many of them signing up for future evaluation studies of the system.

The contribution presented here is concrete in that it advances the current state of the art visual analytics system to provide a method for analyzing players' temporal behaviors unpacking their strategies and problem solving choices over time. The system also allows users to do that at the individual as well as population level, satisfying the users' requirements. However, the system still needs much work.

Besides the interface issues, a known limitation of force-directed layout used by the system is its high running time; a standard implementation can take $O(n^3)$ to complete the placement, with n being the number of nodes [9]. Given less than five seconds to run, the current implementation can handle up to the scale of thousands nodes in a modern browser such as Google Chrome. For higher numbers of nodes (more than tens of thousands nodes), the layout can cause the browser to become unresponsive. To address this issue, one option is to adopt a collapsible hierarchical approach that groups well-connected nodes into clusters represented as compound nodes, thereby reducing the total number of initial nodes to visualize. Compound nodes can be subsequently expanded via user interaction (e.g. mouse clicks).

In order to improve readability in cases of high node numbers, a solution that we are working on is allowing users to generate procedural attraction points that attract and repel nodes in the graph according to user-defined properties. By visually grouping together nodes that share similar properties, these user-defined graph manipulators would greatly improve the legibility of the resulted graph and the possibility to make sense of data, as seen with the system Kinetica [24].

Another direction of improvement is expanding the applicability of Glyph to continuous data domains. Currently the system does not natively handle continuous data, as this data form will lead to an infinite number of states or links. A viable way to tackle this problem is projecting continuous data into a finite countable discrete space, using suitable discretization schemes, such as representing location data as region landmarks instead of exact coordinates.

## 9. ACKNOWLEDGMENTS
We would like to thank Shawn Connor, Keith Devlin, Jordan Lynn, Ben Medler, and Elizabeth Rowe for participating in our requirement analysis study of desired data visualization tools, and Keith Devlin and Randy Weiner from BrainQuake for providing us access to Wuzzit Trouble data. This work is funded by Northeastern University Tier 1 Grant: Learning Analytics.